\documentclass[conference]{IEEEtran}
\IEEEoverridecommandlockouts
\usepackage{cite}
\usepackage{amsmath,amssymb,amsfonts}
\usepackage{algorithmic}
\usepackage{graphicx}
\usepackage{textcomp}
\usepackage{xcolor}
\def\BibTeX{{\rm B\kern-.05em{\sc i\kern-.025em b}\kern-.08em
    T\kern-.1667em\lower.7ex\hbox{E}\kern-.125emX}}

\usepackage[normalem]{ulem}

\usepackage[ruled,vlined,linesnumbered,resetcount]{algorithm2e}
\usepackage{physics}
\usepackage{array}
\newcolumntype{?}{!{\vrule width 1pt}}

\usepackage{url}
\usepackage{breakurl}

\newcommand{\psc}{Pauli string simulation circuit}
\newcommand{\ps}{Pauli string}

\begin{document}

\title{Software-Hardware Co-Optimization for Computational Chemistry on Superconducting Quantum Processors\\

}

\author{\IEEEauthorblockN{Gushu Li}
\IEEEauthorblockA{
\textit{University of California}\\
Santa Barbara, CA \\
gushuli@ece.ucsb.edu}
\and
\IEEEauthorblockN{Yunong Shi}
\IEEEauthorblockA{
\textit{Amazon Braket}\\
New York, NY \\
}
\IEEEauthorblockA{
\textit{University of Chicago}\\
Chicago, IL \\
shiyunon@amazon.com}
\and
\IEEEauthorblockN{Ali Javadi-Abhari}
\IEEEauthorblockA{\textit{IBM Quantum, T. J. Watson Research Center} \\
Yorktown Heights, NY \\
ali.javadi@ibm.com}

}

\maketitle

\begin{abstract}


Computational chemistry is the leading application to demonstrate the advantage of quantum computing in the near term. However, large-scale simulation of chemical systems on quantum computers is currently hindered due to a mismatch between the computational resource needs of the program and those available in today's technology. In this paper we argue that significant new optimizations can be discovered by co-designing the application, compiler, and hardware. We show that multiple optimization objectives can be coordinated through the key abstraction layer of {\em Pauli strings}, which are the basic building blocks of computational chemistry programs. 
In particular, we leverage Pauli strings to identify critical program components that can be used to compress program size with minimal loss of accuracy. We also leverage the structure of Pauli string simulation circuits to tailor a novel hardware architecture and compiler, 
leading to significant execution overhead reduction by up to 99\%. While exploiting the high-level domain knowledge reveals significant optimization opportunities, our hardware/software framework is not tied to a particular program instance and can accommodate the full family of computational chemistry problems with such structure. We believe the co-design lessons of this study can be extended to other domains and hardware technologies to hasten the onset of quantum advantage.

\end{abstract}


\section{Introduction}

Computational chemistry is an important domain in scientific computing that
employs computer simulation to help understand and predict the properties of chemical systems like molecules~\cite{jensen2017introduction}.
It has broad applications in chemistry~\cite{aspuru2018matter}, biology~\cite{reiher2017elucidating}, and material science~\cite{babbush2018low}.
However, simulations of large chemical systems quickly become intractable as the laws governing them lead to equations too complicated to solve efficiently on classical computers~\cite{dirac1929quantum}. For example, more than 1 million node-hours on the Summit supercomputer were recently allocated to chemistry and materials simulation~\cite{olcf}.

Fortunately, quantum computers are naturally suited to solve problems in computational chemistry. In fact, this was the original motivation for Feynman's proposal to build a quantum computer~\cite{feynman1982simulating}.
A leading algorithm for this task is known as the Variational Quantum Eigensolver~(VQE), which has relatively modest requirements in terms of number of qubits and depth of computation, and shows some robustness to errors, all favorable properties for near-term quantum computing~\cite{peruzzo2014variational,mcclean2016theory}.
Small-size molecular simulations using VQE have been experimentally demonstrated with superconducting quantum circuits~\cite{o2016scalable,kandala2017hardware,colless2018computation,google2020hartree} and other technologies~\cite{shen2017quantum,hempel2018quantum,kokail2019self,nam2020ground}.


Despite the recent progress, larger-scale chemistry simulations are not yet feasible on quantum devices. We argue that this is primarily due to three shortcomings in current quantum computing technologies:
1) large program (circuit) size, 2) inefficient hardware architecture, and 3) deficient compiler optimizations.
Each piece is under active research, but rarely in a collaborative way, leading to insufficient overall improvement. To the best of our knowledge, there is no existing united co-optimization solution throughout the application, hardware, and compiler stacks in quantum computing. In this paper we make the case that co-optimizing all three of them can dramatically optimize the overall execution, allowing quantum applications to scale much sooner.

\begin{figure*}
    \centering
    \includegraphics[width=\textwidth]{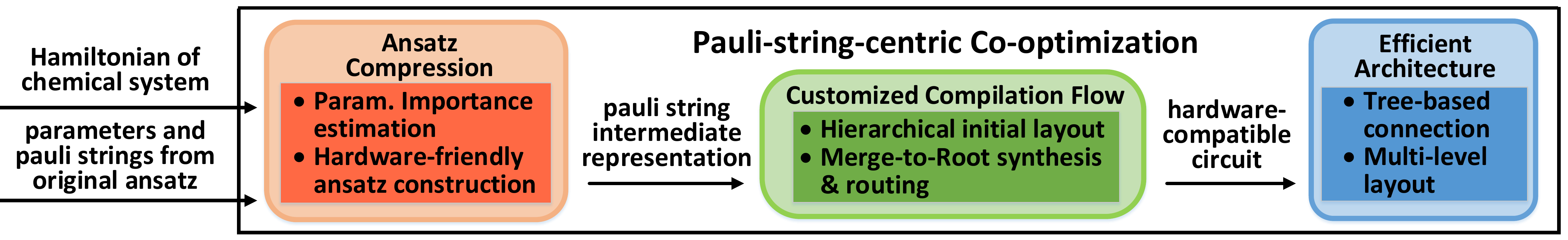}
    \caption{Overview of the proposed Pauli-string-centric software-hardware co-optimization}
    \label{fig:overview}
\end{figure*}

While the co-design principle has been shown to be effective~\cite{staunstrup2013hardware}, it is challenging as the design objectives of different technology stacks may contradict each other.
We briefly review some of these challenges below.

{\bf Application:} Optimizations to reduce the size of VQE circuits have been mostly done theoretically, ignoring the actual execution on the underlying hardware~\cite{lee2018generalized,grimsley2019adaptive,dallaire2019low,ryabinkin2018qubit,ryabinkin2020iterative,tang2019qubit}. VQE is an iterative optimization algorithm and more parameters to optimize over can result in better accuracy. However, this adversely leads to larger circuits and longer time to converge, both undesirable on near-term quantum hardware~\cite{grimsley2019adaptive}. Making the program hardware-friendly~\cite{kandala2017hardware} without keeping its general chemistry structure could prevent it from converging to the right solution effectively~\cite{mcclean2018barren}.

{\bf Hardware architecture:} The quality of superconducting quantum processors has steadily improved in the past few years, while the progress is usually measured by metrics that are oblivious to application performance~\cite{koch2007charge, mckay2016universal, kelly2015state, rosenblatt2019laser, cross2019validating,jurcevic2021demonstration}. Applications generally require high qubit connectivity, but this will cause adverse crosstalk and low yield during device fabrication~\cite{brink2018device,chamberland2020topological,murali2020software}. Making connections sparse will lead to high qubit mapping overhead during application execution~\cite{murali2019full}.

{\bf Compiler optimizations:} State-of-the-art quantum compilers~\cite{Qiskit,quilc,sivarajah2020t} mostly perform optimizations at the gate level where it is easier to reason about program optimization~\cite{soeken2013white,nam2018automated,maslov2008quantum,murali2019noise}, but they miss a large optimization space when compiling VQE programs because they do not exploit the synergy of domain knowledge and hardware information.

In this paper, we co-optimize the algorithm, hardware, and compiler for VQE on superconducting quantum processors through a key observation that \textit{optimizations at different technology stacks can be coordinated through \textbf{\ps s} and their simulation circuits}.
\ps s arise naturally as fundamental building blocks in quantum chemistry simulation. Their unique semantics and structure can be carried through the stack to guide all aspects of the design.
At the algorithm level, the VQE program is dominated by \psc s. The molecule's Hamiltonian (energy operator to be estimated) is also represented by an array of weighted \ps s.
We find that the geometrical interpretation of \ps s can effectively compress the VQE circuit to estimate the same solution with much lower cost.
At the hardware level, the gate pattern of \psc s makes it possible to efficiently support their execution with very few on-chip connections.
Moreover, \psc~synthesis is flexible, allowing us to tailor the compilation flow when deploying VQE programs to the underlying hardware. 
Such property makes it possible to achieve very low execution overhead even on a sparsely-connected hardware architecture.

Our Pauli-string-centric software-hardware co-optimization is shown in Figure \ref{fig:overview}.
\textbf{First}, we introduce a novel VQE circuit compression strategy that takes the Hamiltonian of the target chemical system as an additional input.
The impor-tance of each parameter in the VQE circuit is estimated by \textit{comparing the \ps s} of the circuit with the target Hamiltonian.
Only those parameters that are expected to signi-ficantly affect the final result are kept in a hardware-friendly order. The output of this step is an array of Pauli strings and their parameters, which can be considered as a new interme-diate representation (IR) above quantum circuits.
\textbf{Second}, we design an \textit{X-Tree} superconducting quantum processor archi-tecture that is extremely sparse as it uses the minimal number of physical connections. The sparsity significantly boosts the processor reliability and yield rate. Yet, it does not sacrifice performance
as the connectivity structure is well-suited for the structure of \psc s that appear in various chemistry and physics applications.
\textbf{Third}, we pro-pose a new compilation flow that converts the Pauli string IR directly into executable quantum circuits for the X-Tree ar-chitecture. We determine qubit layouts directly from the Pauli strings (termed \textit{hierarchical initial layout}). We also perform synthesis and mapping in one step (termed \textit{Merge-to-Root}). We show that relying on this higher-level IR, our compiler can map the program to hardware with negligible overhead, as it can adaptively synthesize {\em each} Pauli string according to the current mapping and the underlying X-Tree architecture.

Our co-designed stack is not limited in programmability and can accommodate a wide range of problems in chemistry and physics that are naturally represented by Pauli strings. We show a comprehensive evaluation by simulating various molecules of different sizes and structures. 
Results show that our co-optimization outperforms conventional VQE setups with significant program size reduction, faster convergence speed, mild simulation accuracy loss, more efficient hardware design, and negligible compilation mapping overhead.

Our key contributions can be summarized as follows:
\begin{itemize}
    \item We discover a Pauli-string-centric co-optimization opportunity that can 
    broadly advance variational quantum chemistry simulation of various chemical systems on superconducting quantum processors.
    \vspace{10pt}
    \item We propose three novel optimizations for VQE algorithms, quantum compilers, and superconducting hardware architectures, respectively. Each of them not only focuses on the design objectives of one individual technology but also considers the optimizations in other system stacks. 
    \vspace{10pt}
    \item Our experiments show that our approach outperforms conventional setups of VQE on superconducting quantum processors across a wide range of criteria from software to hardware. 
    On average for nine molecules, when using a 50\% parameter compression ratio, our technique can achieve about  $2.5\times$ convergence speedup and only incur $0.05\%$ error in the simulated energy. It also achieves $99.7\%$ mapping overhead reduction on an optimized architecture with $8\times$ fabrication yield improvement.
    \vspace{5pt}
\end{itemize}

\section{Background}
In this section, we introduce the necessary background to help understand the proposed co-optimization. 
We do not cover basic concepts in quantum computing like qubits, common gates, measurement, and quantum circuits.
We refer the reader to excellent resources such as~\cite{nielsen2010quantum} for more details. 



\subsection{Pauli String and Its Time Evolution Circuit}\label{sec:paulisynthesis}
The central building blocks of chemistry simulation circuits are Pauli string operators.
An $n$-qubit Pauli string $P$ is an array  $P=G_{n-1}G_{n-2}\dots G_{0}$ where $G_{i}\in 
\{I, X, Y, Z\}$ for the $i$th qubit and $0\leq i < n$. $X$, $Y$, $Z$ are the three Pauli operators and $I$ is the identity operator.

\textbf{Time evolution:}
In quantum physics, the time evolution of a system is determined by the system Hamiltonian $H$, and the unitary that represents this time evolution is $exp(i\theta H)$ where $\theta$ is a parameter to represent time.
Usually, we do not directly implement $exp(i\theta H)$ in a quantum circuit since it is hard to directly synthesize $exp(i\theta H)$ into basic single-qubit and two-qubit gates efficiently.
Instead, we first decompose $H$ into a weighted sum of \ps s, i.e., $H=\sum_{j}w_{j}P_{j}$ where $P_{j}$ is a \ps~and $w_{j}\in \mathbb{R}$ is its weight.
The time evolution of \ps s $exp(i\theta P_{i})$ can be easily synthesized. 

\begin{figure}[t]
    \centering
    \includegraphics[width=\columnwidth]{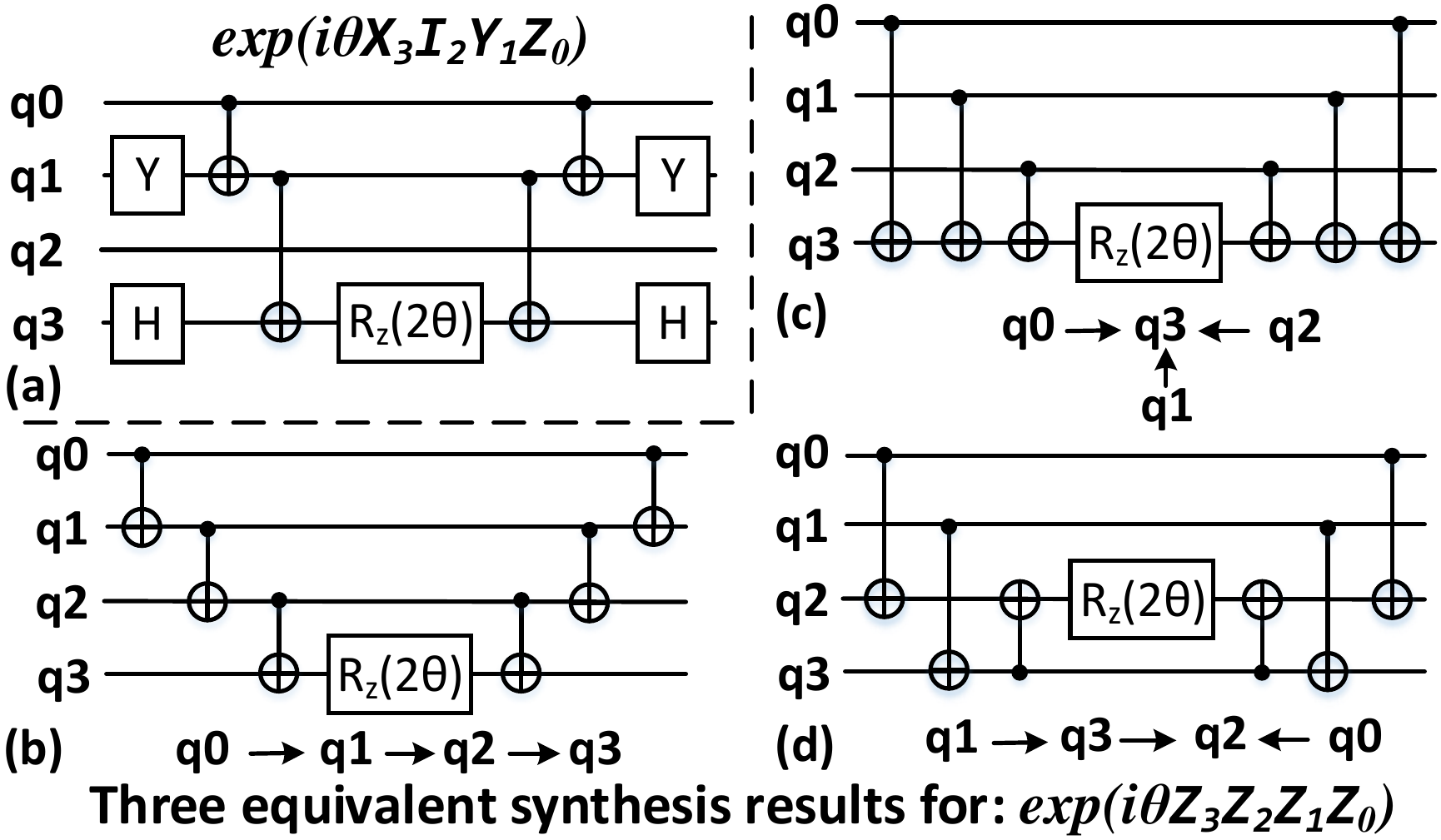}
    \caption{\psc~synthesis examples}
    \label{fig:paulisynthesis}
\end{figure}

\textbf{\psc:}
We introduce the synthesis of \psc s with the examples in Figure~\ref{fig:paulisynthesis}.
Suppose the Pauli string is $XIYZ$ on four qubits and the time parameter is $\theta$.
The circuit in Figure~\ref{fig:paulisynthesis} (a) shows the synthesis result.
The first layer consists of some single-qubit gates.
The rule is that if the operator on one qubit is $X$ (e.g., q3), then we apply an $H$ (Hadamard) gate.
If the operator on one qubit is $Y$ (e.g., q1),  then we apply a $Y$ gate.
If it is $I$ (e.g., q2) or $Z$ (e.g., q0), no single-qubit gate is required.
After applying the single-qubit gates, several CNOT gates will connect all qubits whose corresponding operators are not $I$ in the \ps. In this example, the CNOT gates connect q0, q1, q3 because the operator on q2 is $I$.
We can first connect q0 and q1 and then connect q1 and q3, as shown in Figure~\ref{fig:paulisynthesis}.
Then, a rotation gate is applied to rotate angle $2\theta$ along the $Z$ axis on the last qubit in the CNOT connections (i.e., q3).
Finally, the CNOT gates and single-qubit gates are applied again in the reverse order.
In summary, the \ps~will determine the outermost single-qubit gates and the CNOT gates.
The parameter will only affect the rotation angle of $Z$-rotation gate in the middle.

\textbf{Flexible synthesis:}
The most expensive components for executing a \psc~on a near-term superconducting quantum processor are the CNOT gates before and after the middle rotation gate.
Across various qubit technologies today, (non-local) CNOT gates have an order of magnitude larger latency and error compared to (local) single-qubit gates.
During the synthesis of a \psc, there is flexibility in the pattern of CNOT gates used.
For example, the three circuits in Figure~\ref{fig:paulisynthesis} (b)(c)(d) show three equivalent synthesis result variants of $exp(i\theta ZZZZ)$.
The requirement of the CNOT gates is that they must be connected in a \textit{tree structure} and the CNOT gates are then applied from the leaves to the root (the center rotation gate is applied on the root qubit).
The tree structures of the three variants are shown below the corresponding circuit examples. 
Each qubit is a node and the CNOT gates are represented by directed edges connecting the nodes.
We leverage this flexibility to guide our hardware architecture design and compiler optimizations.

\begin{figure*}
    \centering
    \includegraphics[width=0.98\textwidth]{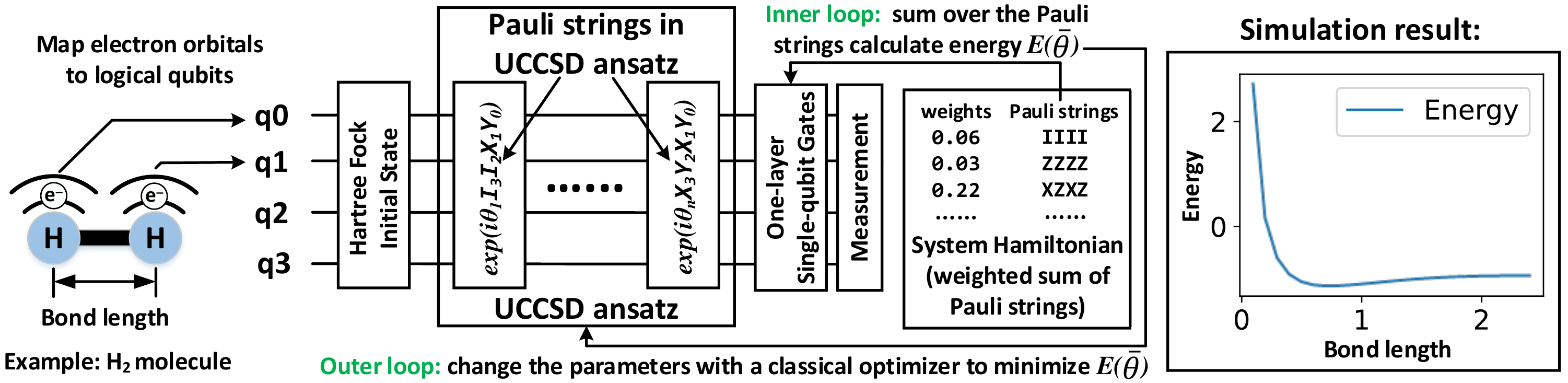}
    \caption{Example of variational quantum chemistry simulation flow and result}
    \label{fig:vqeexample}
\end{figure*}

\subsection{Variational Quantum Computational Chemistry}
We use the example in Figure~\ref{fig:vqeexample} to briefly introduce the basics of VQE algorithm for chemistry simulation.
We recommend~\cite{mcardle2020quantum} for further details.

\subsubsection{Problem encoding}

The first step is to encode the simulation problem, for example a Hydrogen ($\rm H_2$) molecule at a specific bond length (on the left of Figure~\ref{fig:vqeexample}).
To simulate the state of the electrons, we map four candidate orbitals (basis states) that an electron may occupy, and then obtain the system Hamiltonian through standard chemistry tools like PySCF~\cite{PYSCF}. This step is not the focus of our work.

\subsubsection{Circuit construction}\label{sec:vqestructure}
After we map the orbitals to qubits, we need to construct a circuit that can generate a state to represent how the electrons occupy the orbitals.
Figure~\ref{fig:vqeexample} shows the overall structure of this circuit.
After all the qubits are initialized to $\ket{0}$ state at the beginning, the first part on the left is a shallow circuit (applying $X$ gates on some qubits) to prepare an initial state. We use the default Hartree-Fock initial state~\cite{helgaker2014molecular}.
On the right is one layer of single-qubit gates to change the basis prior to measurement, based on the different terms present in the target molecule's Hamiltonian.
These two components only make up a small portion of the entire simulation circuit.
In this work, we focus on the middle part of the circuit: the parameterized state preparation circuit which is known as \textit{ansatz} in the quantum computational chemistry.
The parameters of this circuit are what get optimized during execution.
\textit{This ansatz part makes up the vast majority of the quantum subroutine and is the target of our co-optimization}. 




\subsubsection{Execution flow}
The execution flow of VQE has two major loops. 
For a given set of parameters (denoted by $\bar{\theta}$), we first execute the circuit to generate state $\ket{\psi(\bar{\theta})}$.
Then we measure the expectation value $\bra{\psi(\bar{\theta})}P_{i}\ket{\psi(\bar{\theta})}$ where $P_{i}$ is a \ps~in the decomposition of $H$. 
We iterate over all $P_{i}$s in $H$ to obtain $E(\bar{\theta}) = \sum_{i}w_{i}\bra{\psi(\bar{\theta})}P_{i}\ket{\psi(\bar{\theta})}$.
Changing to measuring different $P_{i}$s only needs to change the last layer of single-qubit gates and the parameters are not changed in this inner loop in Figure~\ref{fig:vqeexample}.
After $E(\bar{\theta})$ is obtained, a classical optimizer will change the parameters $\bar{\theta}$ to minimize $E(\bar{\theta})$.
This optimization may take many steps to converge and this is the outer loop in Figure~\ref{fig:vqeexample}. In this paper, we optimize both the inner loop and outer loop: we discard less important parameters 
for faster convergence and reduce the circuit cost at each iteration by focusing on important sub-circuits and better mapping.
Finally, we obtain a minimal energy $E(\bar{\theta})$ (representing the ground state) of the $\rm H_2$ molecule under the specified bond length.
In a typical simulation task, we will simulate different bond lengths and record ground state energies for these different configurations.

\subsubsection{Simulation result interpretation}
The result of the $\rm H_2$ simulation is on the right of Figure~\ref{fig:vqeexample}.
The X- and Y-axis represent the bond lengths and the simulated ground state energies, respectively.
The minimal simulated ground state energy is achieved when the bond length is around 0.7 \AA~($1$\AA$=10^{-10}m$ and we sample the bond length every 0.1\AA).
The actual bond length measured by physical experiments is $0.74$\AA, which is consistent with the simulation result.

\subsection{UCCSD Ansatz}
The widely-used UCCSD (Unitary Coupled Cluster Singles and Doubles), a chemistry-inspired ansatz~\cite{uccsd1, uccsd2}, is the `standard' ansatz for variational chemistry simulation. 
The terms in a UCCSD ansatz are similar to those in the Hamilto-nian of a chemical system.
Therefore, it is expected that tuning the parameters in UCCSD can make a `good' guess about the ground state.
A UCCSD ansatz of $n$ qubit has $O(n^4)$ parameters and each parameter corresponds to some \ps s.
When implementing UCCSD in a circuit, it becomes a series of \psc s with parameters, as shown in the middle of Figure~\ref{fig:vqeexample}.
Implementing a UCCSD ansatz is very expensive on a superconducting quantum processor due to its large number of parameters and CNOT gates in the synthesized circuit.  
Our techniques will tailor the ansatz, architecture, and compiler for each other to significantly reduce the cost.

\section{Ansatz Compression}\label{sec:ansatzcompression}

To enable chemistry simulation of larger size problems, we first propose to optimize the simulation program at the algorithm level.
We will focus on optimizing the parameterized ansatz because it makes up most of the program. 
The objectives of the ansatz optimization are summarized as follows:
\begin{itemize}
    \item \textbf{Small:} 
    The constructed ansatz should have a small size, i.e., fewer parameters and gates, for shorter execution time and higher fidelity on near-term devices. 
    \item \textbf{Accurate:} The simulation accuracy should not degrade too much  using a smaller
    ansatz 
    with fewer parameters. 
    \item \textbf{Hardware friendly:} The generated ansatz can be mapped onto the target hardware  without too much overhead.
\end{itemize}
Our optimization will start from the UCCSD ansatz, the well-accepted standard ansatz with a large number of parameters ($O(n^4)$ parameters for $n$ qubits).
We seek to eliminate those parameters that contribute the least to final measurement results. 
Doing this precisely for each parameter can be very complex.
Fortunately, in variational algorithms, we do not have to be very precise, as long as the optimization can converge in a reasonable amount of time. The key is to have enough parameters to explore the optimization space and move towards the answer by adjusting those parameters at each iteration. Thus, we only need to {\em estimate} whether a parameter is more likely or less likely to affect the final measurement results.
The ansatz can be compressed by only selecting those circuit components with critical parameters.
The effectiveness of our parameter importance estimation method can be empirically verified later. 
In the rest of this section, we first study how to estimate the importance of each parameter in the UCCSD ansatz. Then we introduce how to construct the ansatz in a hardware-efficient manner.

\subsection{Parameter Importance Estimation}

\begin{figure}[t]
    \centering
    \includegraphics[width=\columnwidth]{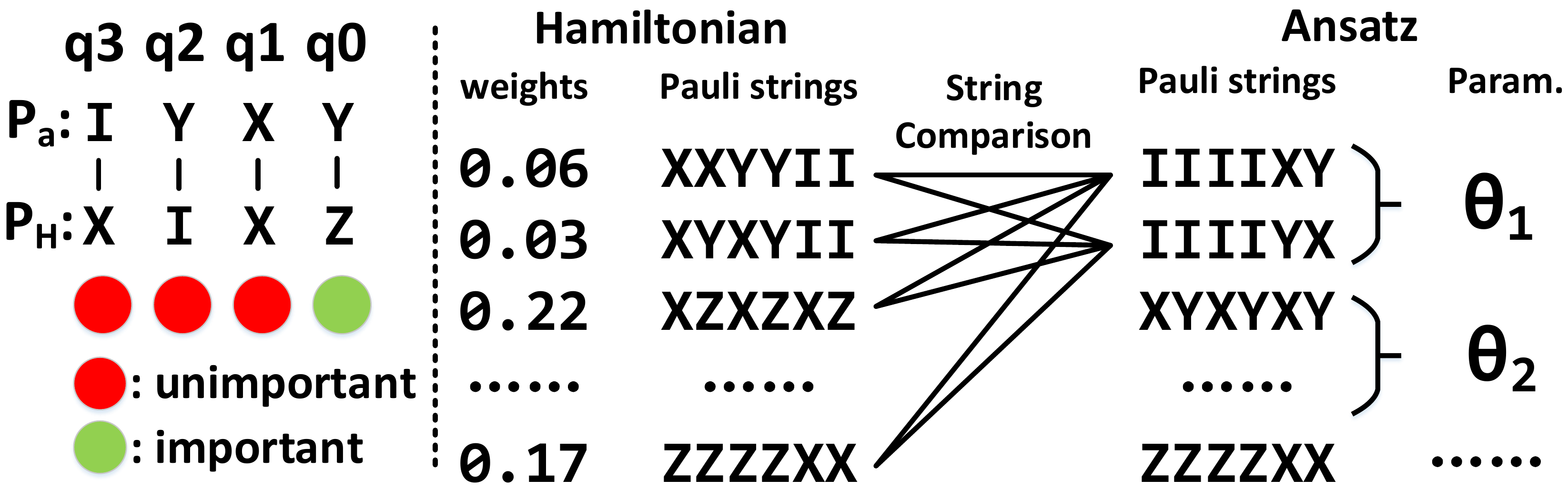}
    \caption{Importance estimation example}
    \label{fig:importanceestimation}
\end{figure}

\begin{figure}[t]
    \centering
    \includegraphics[width=\columnwidth]{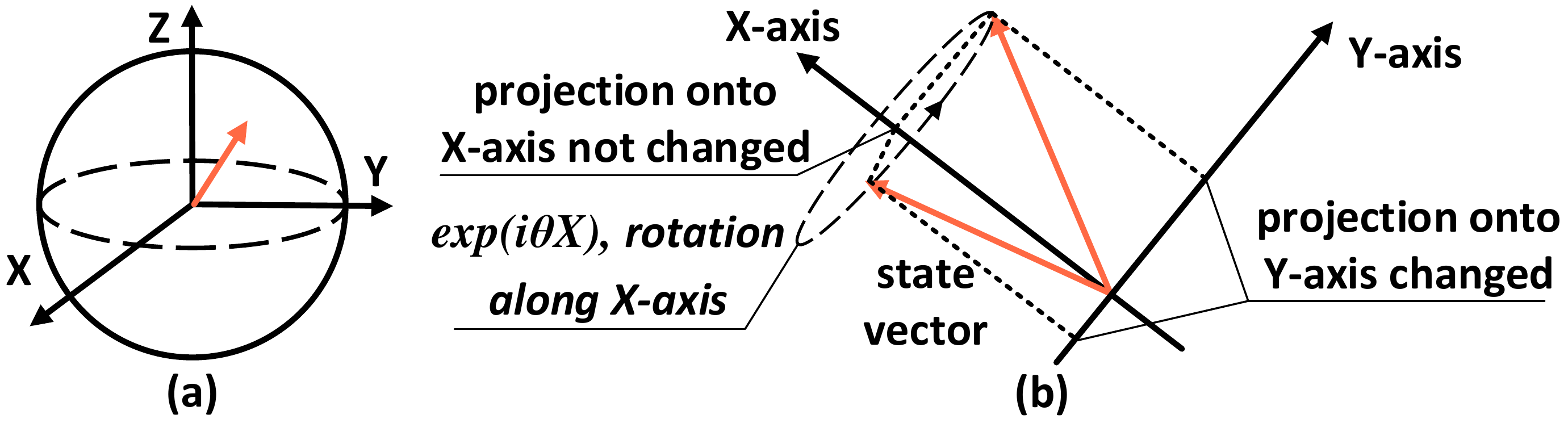}
    \caption{(a) Block sphere with three axes, (b) Effect of state vector rotation}
    \label{fig:geo_interpretation}
\end{figure}

In a VQE simulation, the final observable, which is the Hamiltonian of the target chemical system, is an array of weighted \ps s.
The UCCSD ansatz itself is also an array of \psc s with their corresponding parameters (one parameter can be shared by multiple \ps s).
We first estimate how likely the parameter tuning of each Pauli string in the ansatz can affect the final measurement and then assemble the results to estimate the importance of each parameter. 
The pseudo code of this importance estimation is in Algorithm~\ref{alg:findcriticalcomponents}.
For a given \ps~(denoted by $P_{a}$) in the ansatz, we compare it with each \ps~~(denoted by $P_{H}$) in the Hamiltonian. 
We explain the \ps~comparison method with an example of $P_{a}$ and $P_{H}$ shown on the left of Figure~\ref{fig:importanceestimation}.
For the two Pauli operators on the same qubit $q_{s}$ in the two \ps s being compared, we have the following three cases that will make $P_{a}$ less likely to affect the measurement result of $P_{H}$:

\begin{enumerate}
    \item If the Pauli operator in the $P_{a}$ is `$I$' (e.g., q3), then this \psc~will not apply any gate on $q_{s}$ (as shown in Figure~\ref{fig:paulisynthesis} (a)) and this will make $P_{a}$~less likely to affect the measurement result of $P_{H}$.
    \item If the Pauli operator in the $P_{H}$ is `$I$' (e.g., q2), then when measuring this \psc~in the Hamiltonian, the measurement result on this qubit $q_{s}$ will be always be $1$ and will never be changed with respect to the parameter.
    This makes $P_{H}$ less sensitive to parameter tuning in $P_{a}$.
    \item If the two Pauli operators on $q_{s}$ are the same (e.g., q1), then the effect of changing the parameter in $P_{a}$ will be reduced on the measurement results of $P_{H}$.
    Figure~\ref{fig:geo_interpretation} is a geometrical explanation.
    The state vector of a single qubit can be considered as a unit vector on the Bloch sphere in a three-dimensional  Euclidean vector space (Figure~\ref{fig:geo_interpretation} (a)).
    $X, Y, Z$ can represent three orthogonal axes.
    When applying $exp(-i\theta P)$ ($P\in\{X, Y, Z\}$) on a state vector $\ket{\psi}$, the state vector on the Bloch sphere will rotate around the corresponding axis.
    For example, in Figure~\ref{fig:geo_interpretation} (b), the state is rotating around the $X$-axis after $exp(-i\theta X)$ is applied on it.
    Such rotation will not change the result when we project the state $\ket{\psi}$ onto the same axis, and therefore will not change the measurement result when the observable is $X$.
    
    
    
\end{enumerate}
The only case left is when the two Pauli operators on $q_{s}$ are the different (e.g., q0).
In this case, changing the parameter is very likely to affect the measurement result because rotation along one axis can change the projection onto another axis. For example, in Figure~\ref{fig:geo_interpretation} (b), the projection result on the $Y$ axis is changed after a rotation along the $X$-axis is applied.


Suppose the number of qubits on which the Pauli operators satisfy any of the three conditions above is $d$ and we have $d=3$ in this example.
How likely tuning the parameter of  $P_{a}$ will affect the measurement result of $P_{H}$ is estimated to be the absolute value of the weight of $P_{H}$ multiplied by an exponential decaying term $2^{-d}$.
We repeat this process for all $P_{H}$s in the Hamiltonian and obtain a score of $P_{a}$ in the ansatz. 
After we obtain the scores of each \ps~in the ansatz, the importance of each parameter equals the sum of the scores of all that parameter's corresponding \ps s (note that one parameter can be shared among multiple \ps s).
For example, the importance of $\theta_{1}$ in the example ansatz on the right of Figure~\ref{fig:importanceestimation} is the sum of the scores of the first two \ps s ($IIIIXY$ and $IIIIYX$).
The importance of the rest parameters can be calculated similarly.
The time complexity of our ansatz compression algorithm is $O(n\#(P_a)\#(P_H))$ where $ \#(P_a)$ and $\#(P_H)$ are the numbers of $P_a$s and $P_H$s, respectively, and $n$ is the number of qubits.

\begin{algorithm}[t]
\SetAlgoLined
 
\KwIn{Weighted \ps s of target Hamiltonian $H$, \ps s of one parameter $\theta$}
\KwOut{Importance score of parameter $\theta$}
$importance\_score = 0$\;
\For{$P_{a}$ in all Pauli Strings of parameter $\theta$}{
    \For{$P_{H}$ in all Pauli Strings in $H$}{
    Obtain the importance decay factor $d$ by comparing $P_{a}$  and $P_{H}$\;
    $score$ += $2^{-d}\times$  abs(weight of $P_{H}$)\;
    }
}
\caption{Parameter Importance Estimation}
\label{alg:findcriticalcomponents}
\end{algorithm}

\subsection{Hardware-friendly Ansatz Construction}\label{sec:hardwarefriendlyansatz}

After the importance of each parameter is determined, we can construct the new ansatz and achieve the three objectives mentioned above.
\textbf{First}, since a small size with fewer parameters and \psc s is expected, we will select only part of the parameters and \psc s from the original UCCSD. 
The size of the constructed ansatz can be determined by a given compression ratio.
\textbf{Second}, simulation accuracy is also desired. Therefore, we will select those components that are estimated to be more important than the remaining components.
Changing the parameters in these important components is expected to have a large impact on the final simulated energy. Thus, a lower simulated ground state energy, which will be closer to the true ground state energy, is more likely to be achieved.
For a given compression ratio $\alpha$, if the total number of parameters in the original UCCSD is $K$, then we will select the top $\left \lceil{\alpha K}\right \rceil$ parameters and employ their corresponding \psc s.
\textbf{Third}, we will make the constructed ansatz hardware friendly by putting the \ps s in an \textit{importance-decreasing} order.
Such an order will reduce the overhead when mapping to the target hardware by the compiler because this approach can improve qubit locality in the generated ansatz as explained in the next paragraph.

\textbf{Improving locality:} 
The term qubit locality (similar to data locality in classical computing) in this paper is that the CNOT gates are applied more frequently on some logical qubits in a period of time. 
In quantum chemistry simulation, each qubit represents an orbital but the wavefunction of the electrons is not uniformly distributed on all orbitals.
Different orbitals represent the states with different energies and the electrons are more likely to occupy low energy orbitals because the energy minimum represents a stable ground state.
Therefore, those \psc s that involve low-energy orbitals are more important because changing their parameters will affect the occupancy of the low-energy orbitals.
In our ansatz construction, those \psc s at the beginning of the program mostly include the qubits representing low-energy orbitals.
And these qubits will be frequently involved in the CNOT gates in these \psc s.
This creates gate locality in our constructed ansatz, which makes it easier to be synthesized and mapped later in our compilation. 

The output of our ansatz compression algorithm is a sequence of Pauli strings and their parameters, rather than a typical quantum circuit.
Later in Section~\ref{sec:compileroptimization}, we will have a customized compilation flow to compile the Pauli string sequence into an executable quantum circuit.

\section{Architecture Design}
After a chemistry simulation program is compressed, 
a quantum hardware platform is required to finally execute the optimized program.
In this section, we propose a new superconducting quantum processor architecture to efficiently support variational quantum chemistry simulation.
We first detail the design objectives and physical constraints. 
Then we introduce a new hardware architecture, namely \textit{X-Tree}, and discuss the reasons why it can support VQE circuits 
with both high performance and high efficiency.

\textbf{Design objectives:}
This architecture should support VQE programs with \textbf{high performance}, which means that the simulation programs can be synthesized into circuits and then mapped onto the proposed architecture with low overhead~(i.e., no or few additional SWAP gates). 
It should have as \textbf{few connections} as possible because more connections will increase the probability of frequency collision, lower the yield rate, and also increase crosstalk error.
The architecture should have \textbf{good programmability}, which means it can support programs from the entire UCCSD simulation family for various target chemistry systems.

\textbf{Device modeling and physical constraints:}
We adopt IBM's fixed-frequency transmon qubit and cross-resonance qubit connections~\cite{brink2018device}.
The following practical physical constraints are considered.
Physical qubits are placed on a planar substrate.
One physical qubit can only connect to a limited number of nearby physical qubits directly via bus resonators.
In this work we allow one qubit to connect to at most four neighbors to increase device reliability, but similar architectures with five or six direct connections per qubit have also been built~\cite{iqx}.

\subsection{X-Tree Architecture}
As introduced in Section~\ref{sec:paulisynthesis}, the CNOT gates in the \psc s form a tree structure. 
Therefore, if the physical qubits are connected in a tree, we can match them to the CNOT gates in the chemistry simulation program.
Based on this observation, we propose an X-Tree superconducting quantum processor architecture after considering the design objective and physical constraints mentioned above.

\begin{figure}[t]
    \centering
    \includegraphics[width=\columnwidth]{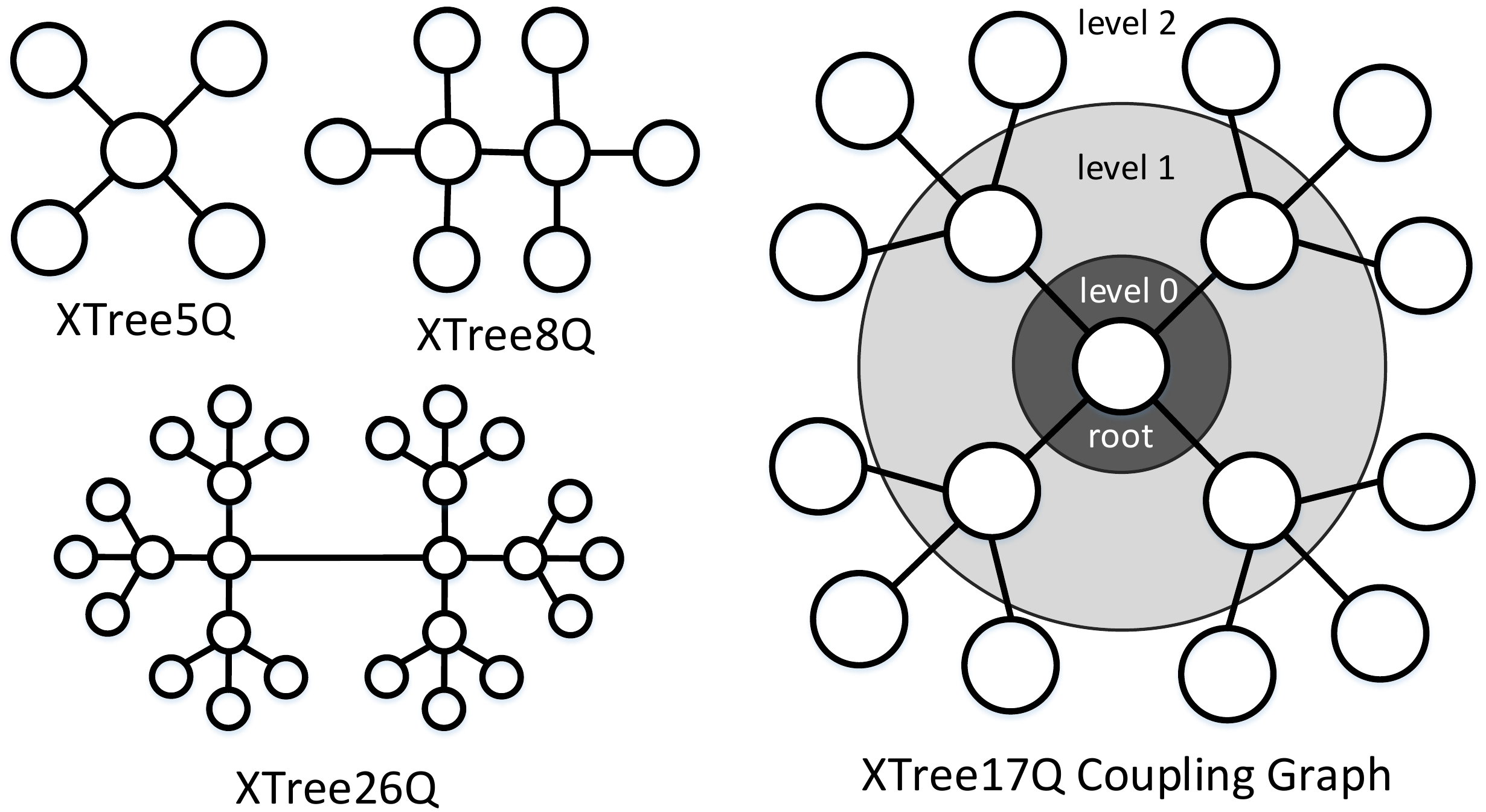}
    \caption{X-Tree architecture examples}
    \label{fig:XTree}
\end{figure}

\textbf{X-Tree architecture construction:}
An X-Tree architecture starts from a \textit{root} qubit. 
Then more qubits are placed and connected.
The key is that the coupling graph formed by the connection is always a tree and there is no circle in the connections.
Figure~\ref{fig:XTree} shows several examples of X-Tree architecture with different numbers of qubits.
We may connect four qubits to the root qubit and obtain the XTree5Q (5-qubit) architecture.
We can add three more qubits to one leaf qubit of XTree5Q to obtain XTree8Q.
Similarly, we can have XTree17Q and XTree26Q architectures by adding more physical qubits. The first generation of IBM's cloud-access quantum computers were compatible with the XTree5Q architecture, but they have since diverged.
Next, we explain why X-Tree architecture can satisfy our three design objectives.

\textbf{Fewer connections:} 
The proposed X-Tree architecture is highly simplified and has only the smallest number of connections ($N-1$ connections for $N$ qubits) to connect all qubits because the coupling graph of an X-Tree architecture is a tree.
As our device yield rate simulations will show, this judicious lowering of connections results in higher \textit{yield rate} in this architecture compared to conventional 2D-grid architectures (which roughly have $2N$ connections for $N$ qubits).
Similarly, \textit{gate crosstalk} errors will be significantly reduced too.

\textbf{High performance and programmability:} We expect the X-Tree architecture to support variational chemistry simulation applications with low mapping overhead since the physical qubit connections naturally fit the logical qubits' CNOT gate connections (both of them are trees).
We also expect programmability since the X-Tree architecture is not tailored to specific any gate-level VQE circuit instances. 
Instead, our design is inspired by the properties of \ps, a high-level algorithm feature, without any assumptions about the simulated system.
However, the physical connection tree is not identical to the CNOT gate connection trees since there are different \ps s on different qubits for different simulation programs.
Compiler optimizations are still required to deploy the chemistry simulation program onto the X-Tree architecture, which will be explained in the next section.

\section{Compiler Optimization}\label{sec:compileroptimization}

Although the X-Tree architecture has been designed to match the tree pattern of gates in a typical quantum chemistry program, we will show that state-of-the-art compilers are not well suited for taking maximum advantage of it. A traditional quantum compilation flow separates high-level synthesis from mapping onto the architecture.
That is, it will first convert the \ps s and the parameters into concrete \psc s using a uniform CNOT synthesis plan. For example, Qiskit~\cite{Qiskit} synthesizes the CNOTs in a \psc~in a chain structure like Figure~\ref{fig:paulisynthesis} (b).
However, recall that there is great flexibility in how each \psc~is synthesized: as long as the non-trivial qubits in the Pauli string are connected by a tree, it does not matter which connections we use. This is the key insight that allows us to adaptively synthesize and map each \psc~in the larger ansatz. The approach taken by previous compilers fails to recognize this flexibility. Once the circuit is synthesized, it is exceedingly hard to find such high-level semantics, and mapping a poorly synthesized program on a sparse architecture can incur a very high cost.

In this section, we introduce the third optimization,
a tailored compiler optimization to efficiently synthesize and map 
variational quantum chemistry simulation 
programs to X-Tree architectures with very low 
overhead. We will show that this tailored approach incurs an overhead of around 99\% lower than a traditional compiler for the same architecture, and even 97.7\% lower than mapping to a dense architecture but without leveraging such compiler optimizations.

Our compiler optimization performs circuit synthesis and qubit mapping collaboratively in two steps. 
First, we determine an initial qubit layout, based on the ansatz Pauli strings only, before the program is synthesized to gate sequences. 
Then we perform circuit synthesis and qubit routing (inserting SWAPs) simultaneously onto the X-Tree architecture.

\subsection{Hierarchical Initial Layout}

Since the program is not converted to gates yet, our initial qubit layout algorithm will directly analyze the high-level program and provide an initial qubit layout.
This is possible because our proposed X-Tree architecture has different physical qubit levels.
For example, in the XTree17Q architecture in Figure~\ref{fig:XTree}, the center (root) qubit has level $0$ as it is on average closer to all other qubits. The four qubits surrounding the root have level $1$, and the leaves have level $2$.
Similarly, we can also discover different priorities for different logical qubits in a chemistry simulation program. 
The states represented by some orbitals are closer to the true ground state of the electrons, thus the logical qubits corresponding to these orbitals will appear in more \psc s and will participate in more CNOT gates (as discussed in Section~\ref{sec:hardwarefriendlyansatz}). 
We place these logical qubits on lower-level physical qubits, ensuring that they can reach other qubits with shorter paths.

\begin{algorithm}[t]
\SetAlgoLined
 
\KwIn{Pauli strings in the simulation program, an X-Tree architecture with qubits at different levels}
\KwOut{Initial logical-to-physical qubit mapping}
\For{$P_{i}$ in all Pauli Strings}{
    \If{the qubit \textbf{j} and \textbf{k} appear in $P_{i}$}{
    $Mat(j,k) += 1$\;
    }
}
$Qubit\_occurrence = \sum_{k}Mat(j,k)$\;
$Logical\_qubit\_order=ArgSort(Qubit\_occurrence)$\;
\For{qubit j in $Logical\_qubit\_order$}{
Map qubit j to the physical qubit in the lowest available level\;
\If{there are multiple possible parent qubit k}{
select k = argmax(Mat(j,k))
}
}
\caption{Hierarchical Initial Layout}
\label{alg:initiallayout}
\end{algorithm}

\begin{figure}[t]
    \centering
    \includegraphics[width=\columnwidth]{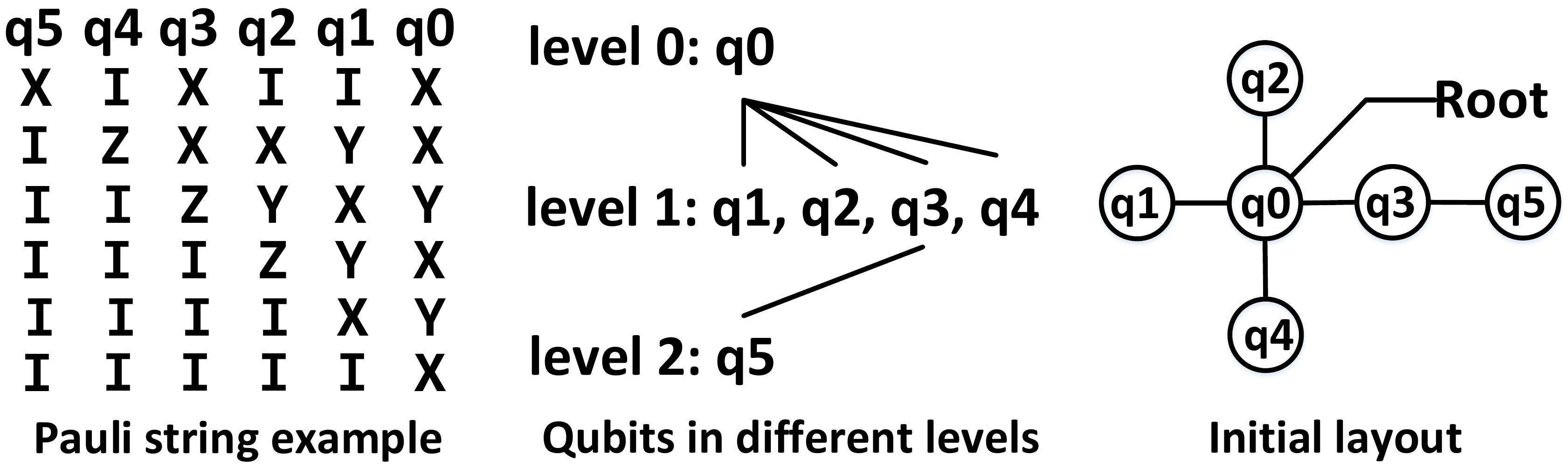}
    \caption{Initial layout example}
    \label{fig:initiallayout}
\end{figure}

Our hierarchical initial layout algorithm is based on such heterogeneity of the logical and physical qubits.
The pseudocode is in Algorithm~\ref{alg:initiallayout} and we explain the algorithm with the example in Figure~\ref{fig:initiallayout}.
We first determine which qubits appear in more \ps s.
A matrix will record the number of instances when qubit $i$ and $j$ appear in the same \ps~(the first loop).
Then we can know which qubits connect to other qubits more by taking summation in one dimension.
Finally, we sort the logical qubits by their connectivity requirements and place them on the X-Tree architecture from level $0$ outwards. 
In Figure~\ref{fig:initiallayout}, we put q0, which appears in all \ps s, on the level 0 root and put q1, q2, q3, q4 in the four level 1 physical qubits.
In case of multiple available spots, we attach to a parent qubit which shares the largest number of common \ps s with the logical qubit to be allocated (the parent is already allocated a physical spot because it is in a lower level).
In the example in Figure~\ref{fig:initiallayout}, q5 has been assigned level $2$ as it participates in only one \ps. Of the qubits it shares a \ps~with, q3 is one level up and so chosen as q5's parent.

\begin{figure}[t]
    \centering
    \includegraphics[width=\columnwidth]{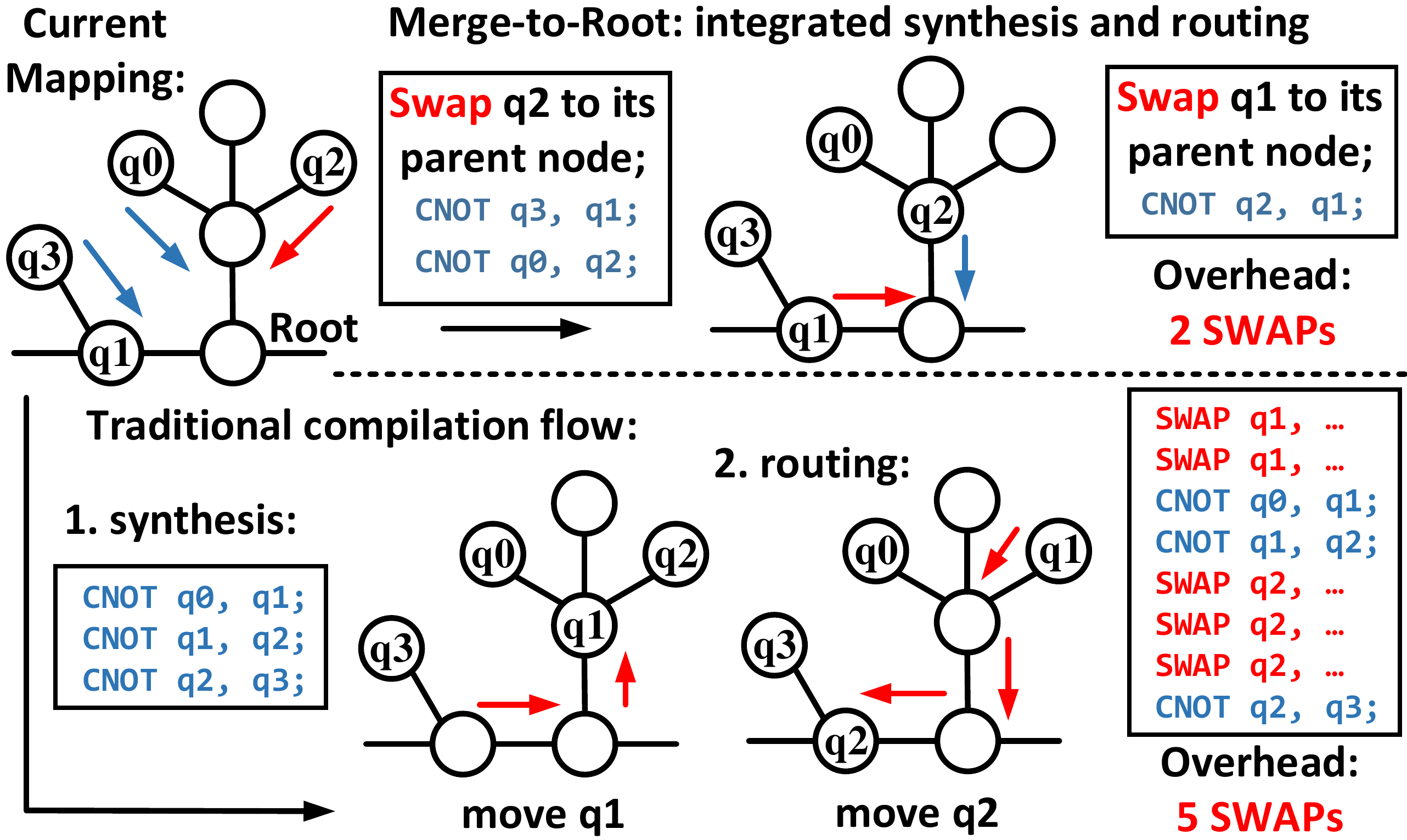}
    \caption{ Merge-to-Root vs traditional compilation
    }
    \label{fig:mergetorootexample}
\end{figure}

\subsection{Merge-to-Root Circuit Synthesis and Qubit Routing}

After the initial qubit mapping is determined, we need to synthesize the \psc s into concrete circuits and resolve all the CNOT gate dependency issues caused by the limited on-chip qubit connection.
We propose a \textit{Merge-to-Root} algorithm to synthesize the simulation circuits and determine how to insert SWAPs for remapping qubits.
For each \psc, the two layers of the single-qubit gates at the beginning and the end are fixed. 
We only need to synthesize two 
CNOT trees and the center rotation gate as introduced in Section~\ref{sec:paulisynthesis}.
The pseudocode is in Algorithm~\ref{alg:mergetoroot} and we explain it with the example in Figure~\ref{fig:mergetorootexample}.

Suppose we need to compile the simulation of \ps~on four logical qubits, q0, q1, q2, and q3. 
Their current mapping on an X-Tree architecture is shown on the top left of Figure~\ref{fig:mergetorootexample}.
Our merge-to-root compilation starts from the outermost physical qubits. 
We can find that q0, q2, and q3 are currently mapped onto level 2 physical qubits.
We check the parent qubits (at level 1) of these outermost qubits.
If a parent qubit is holding a logical qubit in the simulation circuit, e.g., the parent qubit of q3 is the one q1 is mapped onto, then we can synthesize a CNOT between these two qubits.
If not, we will find one qubit in the current level and swap it to this parent qubit.
For example, the parent qubit of q0 and q2 is not in the \ps.
We first select one of them and SWAP it to the parent qubit.
We will select the qubit that will appear more times in the follow-up \ps s.
Suppose we 
move q2 
to the parent physical qubit.
We can now synthesize a CNOT between q0 and q2.
The procedure above synthesizes all CNOTs that are between level 2 and 1 with just one SWAP overhead.
It will be repeated from the outer levels to the inner levels until the last qubit.
For level 1 qubits (q1 and q2 in this example), we can move q1 and then synthesize the last CNOT between q2 and q1.
This synthesis of the left CNOT tree is now completed with only two SWAPs in total.
The center rotation gate can then be applied on q1.
The right CNOT tree can be synthesized similarly in a reversed order from the inner levels to the outer levels.
The time complexity of our compiler optimization algorithm is $O(n\#(P_a))$ where $n$ is the number of qubits and  $\#(P_a)$ is the number of Pauli strings in the ansatz.

\begin{algorithm}[t]
\SetAlgoLined
 
\KwIn{Initial qubit layout, Pauli strings in the simulation program, a X-Tree architecture with qubits of $K$ different levels}
\KwOut{A hardware compatible circuit}
\For{$P_{i}$ in all Pauli Strings}{
\tcp{Synthesize left CNOT tree}
   \For{level $k$ from $K-1$ down to $1$ }{
        \If{a qubit at level $k$ is in $P_{i}$ but its parent qubit $q_{p}$ at level $k-1$ is not in $P_{i}$}{
        select one of  $q_{p}$'s  child qubits that are in $P_{i}$ and SWAP it with $q_{p}$\;
        }
        Synthesize all CNOTs from level $k$ to $k-1$\;
   }
}
Apply the center rotation gate on the last qubit\;
Synthesize the right CNOT tree accordingly\;
\caption{Merge-to-Root Synthesis and Routing}
\label{alg:mergetoroot}
\end{algorithm}


\textbf{Comparing with traditional compilation:} The lower half of Figure~\ref{fig:mergetorootexample} also shows the compilation results of the left CNOT tree from traditional compilation flow.
The left CNOT tree will first be synthesized into three CNOT gates.
Then a mapping algorithm will try to move the qubits to satisfy the dependencies of the three CNOT gates.
In this example, we first move q1 by two SWAP gates to execute the first two CNOT gates.
We then move q2 by three SWAP gates to execute the last CNOT gate.
The total overhead is five SWAPs, which is much higher than that of our Merge-to-Root compilation. The key is that, comparing with traditional compilation, Merge-to-Root will synthesize entirely different CNOTs adapted to the current mapping and the architecture.

\section{Evaluation}

We evaluate the proposed co-optimization with carefully designed experiments over a wide range of chemistry simulation benchmarks to show the improvements from the algorithm, hardware, and compiler levels.

\subsection{Experiment Setup}

\textbf{Benchmarks:} 
We select nine molecules of various sizes and geometrical structures. 
The names of the molecules and the information of their simulation circuits using the original full UCCSD ansatz are listed in Table~\ref{tab:benchmark}.
Note that `\# of Pauli' means the number of \ps s.

\begin{table}[t]
  \centering
  \caption{Benchmark molecules and their original cost}
  \small
   \resizebox{\columnwidth}{!}{ \begin{tabular}{|c|c|c|c|c|}
    \hline
     & \# of Qubits & \# of Pauli & \# of Param. & \# of Gates (CNOTs)  \\
    \hline
    $\rm H_{2}$    &   4    &   12    &  3      &  150 (56) \\
    \hline
    $\rm LiH$   &   6    &   40    &   8    &  610 (280) \\
    \hline
    $\rm NaH$   &   8    &   84    &   15    &  1476 (768) \\
    \hline
    $\rm HF$    &   10    &   144    &   24    & 2856  (1616)  \\
    \hline
    $\rm BeH_{2}$  &  12     &   640    &   92    & 13704  (8064) \\
    \hline
    $\rm H_{2}O$  &   12    &   640    &   92    & 13704 (8064)  \\
    \hline
    $\rm BH_{3}$   & 14  &   1488    &  204     &  34280 (21072)\\
    \hline
    $\rm NH_{3}$   &    14   &   1488    &   204    & 34280 (21072) \\
    \hline
    $\rm CH_{4}$   &    16   &   2688   &   360    &  66312 (42368) \\
    \hline
    \end{tabular}%
    }
  \label{tab:benchmark}%
\end{table}%

\textbf{Metric:}
The simulation accuracy is measured by the simulated ground state energy of the target molecule.
We adopt atomic 
units that are more convenient for computational chemistry.
The energy unit is Hartree
($1 \ \text{Hartree} \approx 4.36 \times 10^{-18} \  \text{Joules}$). 
The bond length unit is Angstrom 
($1 \ \text{Angstrom} = 10^{-10} \ \text{meter}$).
The convergence speed is indicated by the number of iterations in the parameter optimization (outer loop in Figure~\ref{fig:vqeexample}). 
A smaller number of iterations means that the simulation converges faster.
Compiler optimizations are evaluated by the gate count in the post-compilation circuit, a widely used metric in previous studies~\cite{zulehner2018efficient,siraichi2018qubit,li2019tackling}.
A more effective compiler optimization will result in a lower gate count in the post-compilation circuit.
The CNOT 
count 
is of particular interest owing to the much higher error rate and longer latency compared to single-qubit gates.

\textbf{Implementation:}
We implement the proposed optimizations  based on
IBM's Qiskit~\cite{Qiskit} and perform experiments with classical simulators in Qiskit. 
The Hamiltonian of the simulated molecule is generated by PySCF~\cite{PYSCF} with STO-3G orbitals~\cite{hehre1969self} and Jordan-Wigner encoding~\cite{jordan1928paulische}.
We freeze the core electrons and only simulate the interaction of the outermost electrons.
We use the default UCCSD ansatz from Qiskit Aqua library~(version 0.8.0).
The parameters are optimized using the Sequential Least Squares Programming \cite{kraft1988software} solver.
The noise-free simulations are performed with Qiskit Aer statevector simulator  and the noisy simulations are performed with Qiksit Aer qasm simulator (version 0.6.0).
For the hardware yield rate, we adopt the yield simulation method and qubit frequency allocation algorithm in~\cite{li2020towards}.
All experiments are performed on a MacBook Pro with 2.8 GHz Quad-Core Intel Core i7 CPU and 16GB 2133MHz LPDDR3 memory.

\subsection{Experiment Methodology}
\textbf{Baseline:} The software baseline is the original UCCSD ansatz~\cite{peruzzo2014variational}, denoted by `Orig. UCCSD'. The true ground state energies for reference, denoted by `Ground State', are obtained by directly calculating the eigenvalue of the Hamiltonian of the target system.
The hardware baseline is IBM's 17-qubit device (Grid17Q) with a 2D grid connection~\cite{brink2018device} (shown on the left of Figure~\ref{fig:yieldcompare}) for a fair comparison with our 17-qubit X-Tree device (XTree17Q) employing the same number of qubits.
The compiler baseline is SABRE~\cite{li2019tackling} (SAB), a state-of-the-art general-purpose mapping algorithm in Qiskit.

\textbf{Configurations:} We apply the parameter compression method in Section~\ref{sec:ansatzcompression} with five compression ratios: 10\%, 30\%, 50\%, 70\%, 90\%. They are denoted by `10\% Param.' to  `90\% Param.'
We also generated ansatzes by randomly selecting 50\% parameters (denoted by `Rand. 50\%').


\subsection{Simulation Accuracy and Convergence Speedup}\label{sec:accuracyandconvergence}
\begin{figure*}[t]
    \centering
    \includegraphics[width=0.992\textwidth]{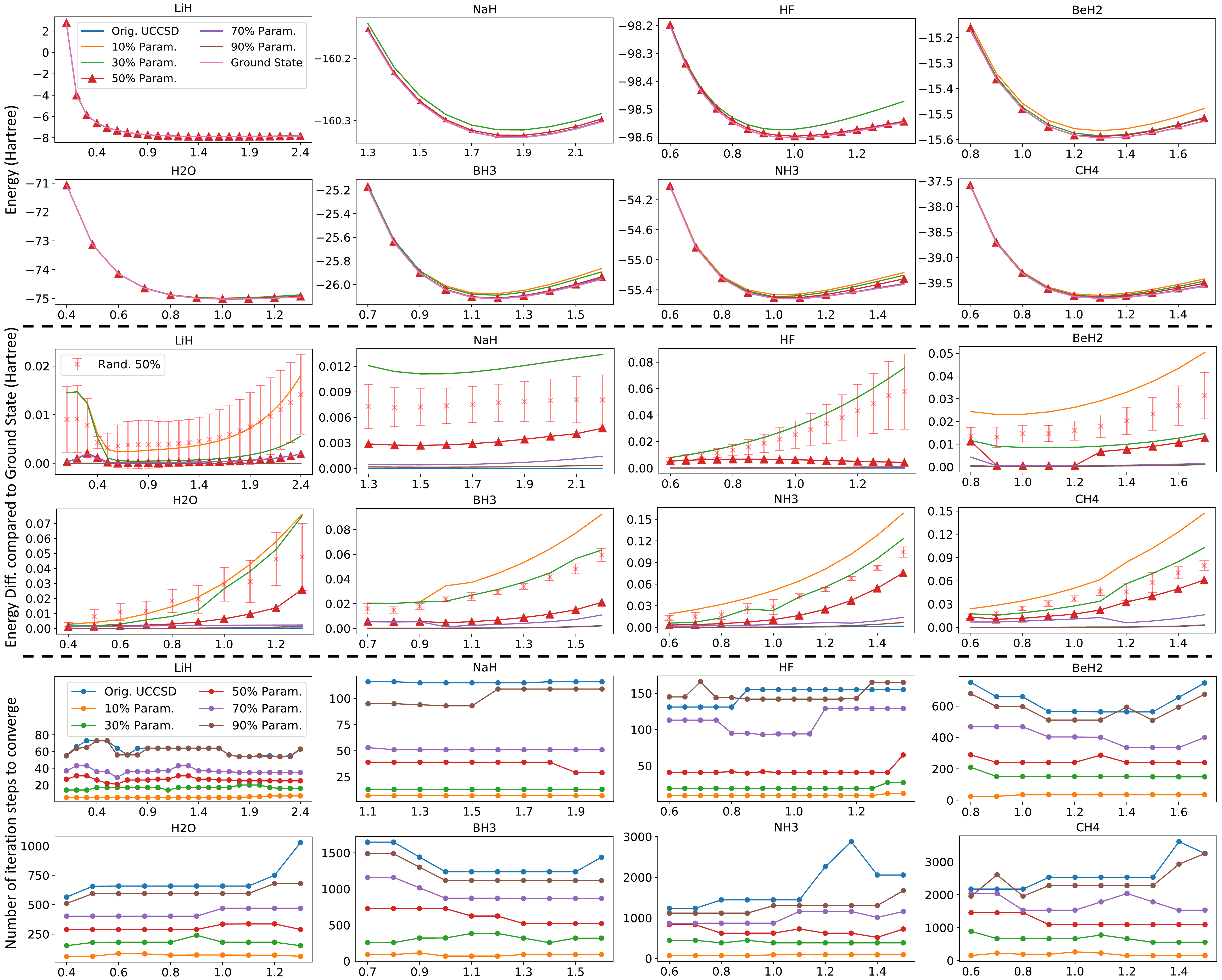}
    \caption{Accuracy and number of iterations vs various parameter reduction ratios}
    \label{fig:accuracy}
\end{figure*}

Figure~\ref{fig:accuracy} shows the simulation accuracy and the convergence speed of our compressed ansatz.
The results of $\rm H_{2}$ is omitted since its circuit is small with only three parameters.
There are three parts in Figure~\ref{fig:accuracy}. The X-axes represent the bond lengths of the simulated molecules.
The Y-axes represent the simulated energy, simulated energy difference, and the number of iterations steps as labeled on the left of Figure~\ref{fig:accuracy}.
The top part shows simulated energies at different bond lengths.
The simulation results of the compressed ansatzes are close to that of the full UCCSD and the true ground states.
The more parameters we keep, the more accurate simulation we can obtain.
To better understand the amount of accuracy loss, the middle part of Figure~\ref{fig:accuracy} shows the energy difference between different experiment configurations and their corresponding true ground states.
For example, the energy differences for `50\% Param.' are usually only at the level of about $0.05\%$.

\begin{figure}[t]
    \centering
    \includegraphics[width=\columnwidth]{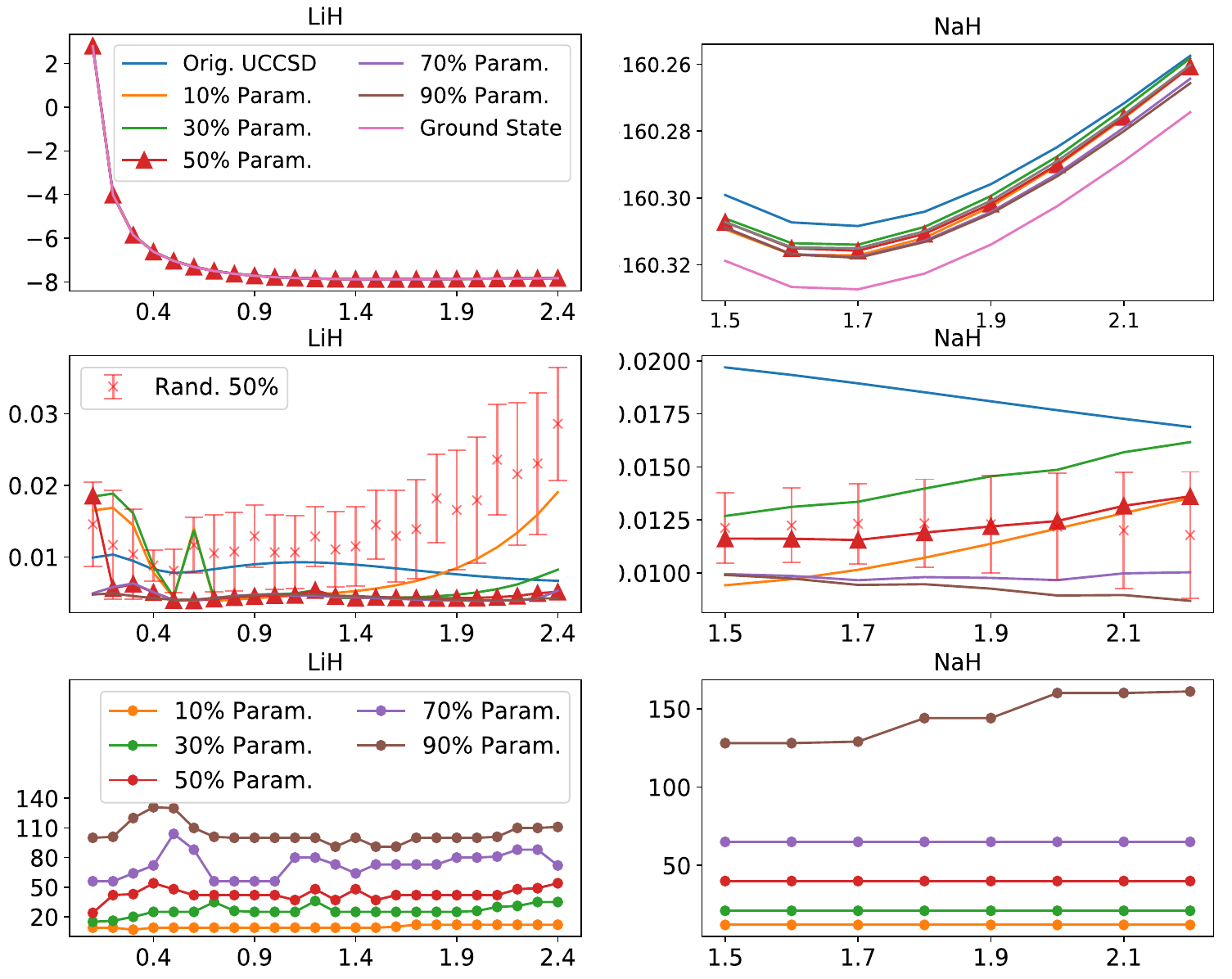}
    \caption{Noisy simulation case studies on $\rm LiH$ and $\rm NaH$  
    }
    \label{fig:noisysim}
\end{figure}
\begin{table*}[t]
  \centering
  \caption{Mapping overhead comparison of different compilation approaches}
  \small
  \resizebox{\textwidth}{!}{
    \begin{tabular}{|c?c|c|c|c|c?c|c|c|c|c?c|c|c|c|c?c|c|c|c|c|}
    \hline
          & \multicolumn{5}{c?}{Original \# of CNOTs} & \multicolumn{5}{c?}{MtR on XTree17Q (\# of CNOTs)}  & \multicolumn{5}{c?}{SAB on XTree17Q (\# of CNOTs)} & \multicolumn{5}{c|}{SAB on Grid17Q (\# of CNOTs)} \\
    \hline
    Ratio      & 10\% & 30\% & 50\% & 70\% & 90\%   & 10\% & 30\% & 50\% & 70\% & 90\%   & 10\% & 30\% & 50\% & 70\% & 90\%   & 10\% & 30\% & 50\% & 70\% & 90\% \\
\hline
$\rm H_2$    & 48    & 48    & 52    & 56    & 56    & 0     & 0     & 0     & 6     & 6     & 0     & 0     & 0     & 0     & 0     & 0     & 0     & 0     & 0     & 0 \\
 \hline
    $\rm LiH$   & 80    & 208   & 256   & 272   & 280   & 0     & 6     & 6     & 12    & 18    & 48    & 126   & 132   & 150   & 168   & 0     & 6     & 9     & 15    & 18 \\
\hline
    $\rm NaH$   & 176   & 448   & 672   & 736   & 764   & 0     & 0     & 0     & 3     & 21    & 162   & 777   & 1002  & 1197  & 1470  & 12    & 12    & 87    & 120   & 123 \\
\hline
    $\rm HF$    & 400   & 912   & 1264  & 1552  & 1608  & 0     & 0     & 0     & 6     & 36    & 633   & 1863  & 2034  & 2163  & 2502  & 87    & 126   & 267   & 372   & 612 \\
\hline
    $\rm BeH_2$  & 1504  & 3808  & 5696  & 7248  & 7984  & 3     & 6     & 24    & 51    & 228   & 3315  & 6513  & 13416 & 14268 & 17862 & 621   & 1395  & 4005  & 5253  & 8091 \\
\hline
    $\rm H_2O$   & 1536  & 3840  & 5712  & 7280  & 7988  & 0     & 12    & 18    & 75    & 135   & 3132  & 7764  & 12495 & 13266 & 15618 & 1110  & 1725  & 2034  & 2514  & 3156 \\
\hline
    $\rm BH_3$  & 3664  & 9632  & 14560 & 18368 & 20824 & 0     & 39    & 108   & 237   & 606   & 9489  & 23811 & 35289 & 45603 & 46395 & 2163  & 7632  & 9654  & 17010 & 21165 \\
\hline
    $\rm NH_3$   & 3680  & 9696  & 14592 & 18480 & 20824 & 0     & 30    & 72    & 183   & 522   & 11646 & 20622 & 35523 & 42348 & 48447 & 1959  & 5844  & 8568  & 12375 & 13668 \\
\hline
    $\rm CH_4$   & 7136  & 19040 & 28992 & 36656 & 41632 & 0     & 45    & 120   & 366   & 1005  & 23796 & 56799 & 79821 & 99831 & 111876 & 4788  & 18939 & 25173 & 33792 & 39729 \\
\hline
    \end{tabular}%
    }
  \label{tab:mappingoverhead}%
\end{table*}%

\textbf{Effective parameter selection:} 
We show the effectiveness of our parameter selection method by comparing the ansatzes generated by our compression method with those
constructed by randomly selected parameters.
For the ansatzes
with 50\% randomly selected parameters, we generate five different random parameter selections for each molecule at each simulated bond length.
The simulation result distribution of `Rand. 50\%' is demonstrated by the mean and standard deviation of the simulated energies.
It can be observed that the `50\% Param.' ansatzes
generated by our optimization outperform the  `Rand. 50\%' with better accuracy and the simulated energies are closer to the true ground state energies.
The accuracy of `Rand. 50\%' is similar to that of `30\% Param.', which means that our optimization can select 30\% of parameters but achieve the same level of  accuracy from randomly selecting 50\% of the parameters.
This comparison proves that our ansatz compression algorithm is very effective.
The execution time of our ansatz compression is negligible compared with the VQE execution itself.
For example, it requires several minutes to compress the ansatz for $\rm CH_{4}$ while it takes over ten CPU hours to simulate $\rm CH_{4}$ with VQE at one bond length.

\textbf{Convergence speedup:}
The bottom part of Figure~\ref{fig:accuracy} shows the number of iterations to converge.
The compressed ansatzes with fewer parameters can converge much faster with smaller numbers of parameter optimization steps.
The numbers of parameter optimization steps are reduced by $14.3\times$, $4.8\times$, $2.5\times$, $1.6\times$, and $1.1\times$ on average for the five parameter compression ratios of 10\% to 90\%, respectively.

There is also a subtle implication in computation reliability when the computation concludes faster. Quantum computers are calibrated to reduce gate errors. After a few hours, the physical properties of the system drift causing the calibration to become stale. At current experimental speeds, a full VQE experiment can easily take hours to converge, which makes these speedups a boost to reliability as well.



\subsection{Noisy Simulation Case Studies}
We study the effect of hardware noise on $\rm LiH$ and $\rm NaH$ as case studies.
Our simulation adopts a depolarizing error model with realistic CNOT error rates of 0.0001~\cite{malekakhlagh2020first}.
Figure~\ref{fig:noisysim} shows the simulation results. Similar to Figure~\ref{fig:accuracy}, the first, second, and third rows are the overall simulated energies, energy differences, and number of iterations, respectively. 
We observe that our compressed VQE can still demonstrate the correct landscapes of the molecule energy under different bond lengths. We can also observe interesting trade-offs between parameter pruning and accuracy in noisy regimes.
For $\rm LiH$, the error first decreases from ‘10\% Param.’ to ‘50\% Param.’ due to the increasing parameters. After that, the error does not change significantly from ‘50\% Param.’ to ‘90\% Param.’ because the effect of more parameters is masked by the increasing gate error. ‘50\% Param.’ is a sweet spot for LiH.
For $\rm NaH$, the balance is different. The error first increases from ‘10\% Param.’ to ‘30\% Param.’ and then drops from ‘30\% Param.’ to ‘90\% Param.’  This suggests that we should either select ‘10\% Param.’ or ‘90\% Param’.
Such trade-offs depend on the molecule Hamiltonian, the bond length configuration, hardware noise strength, and maybe other factors. A comprehensive research into these trade-offs is left as future work.

\subsection{Hardware Efficiency}
\begin{figure}[t]
    \centering
    \includegraphics[width=\columnwidth]{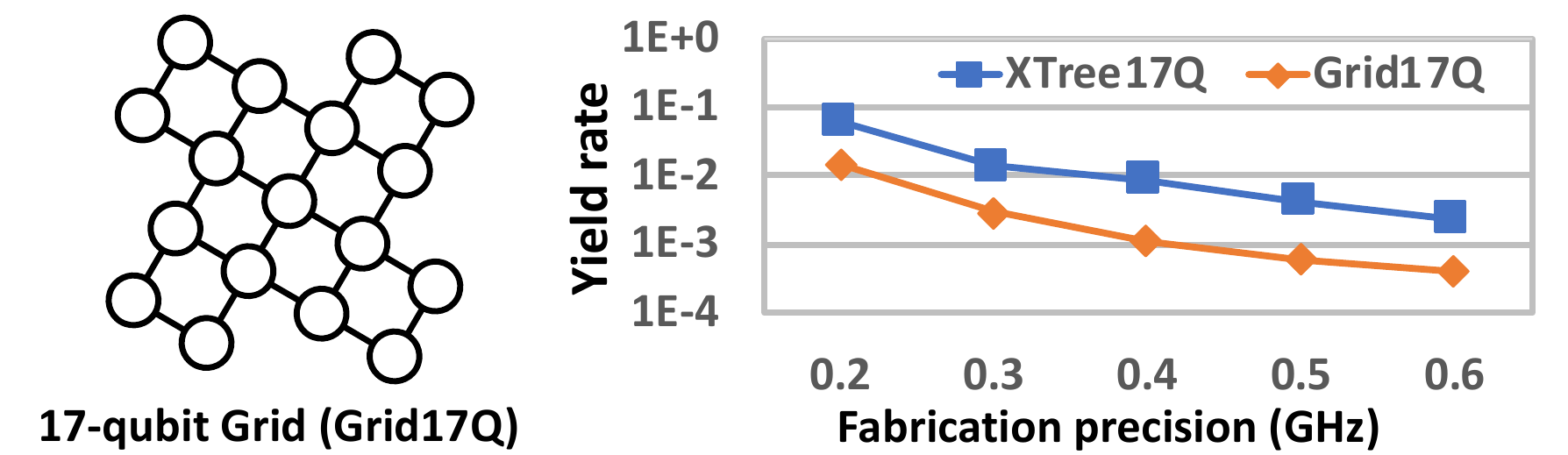}
    \caption{Grid17Q architecture and the yield rate comparison}
    \label{fig:yieldcompare}
\end{figure}

We evaluate our hardware design by comparing the XTree17Q architecture with baseline Grid17Q.
Both of them have 17 physical qubits.
Figure~\ref{fig:yieldcompare} shows the yield rates of XTree17Q and Grid17Q for various fabrication precision parameters from 0.2 GHz to 0.6 GHz.
The yield rate of the XTree17Q architecture is about $8\times$ higher than that of the Grid17Q architecture.
This is because Grid17Q has 24 connections while XTree17Q only employs 16 connections.


\subsection{Mapping Overhead Reduction}
Table~\ref{tab:mappingoverhead} shows the mapping overhead (i.e., the number of additional CNOT gates) of our Merge-to-Root~(MtR) compilation (including our initial layout algorithm) vs. the baseline compilation (SAB) on XTree17Q and Grid17Q architectures.

We first compare MtR on XTree17Q vs. SAB on XTree17Q.
The sparse connectivity of XTree17Q makes the mapping overhead very high for the general-purpose SAB compiler.
The number of additional CNOTs is about $177\%$ of the CNOT count of the original circuits.
This is even worse for larger benchmarks. For $\rm CH_{4}$, the number of additional CNOTs for SAB is about $288\%$ of the original CNOT count.
However, our MtR compilation incurs dramatically smaller overhead. 
For all tested benchmarks, the number of additional CNOTs is on average $1.4\%$ of the original CNOT count.
Therefore, our MtR compilation reduces the mapping overhead to only about $\bf 1\%$ of the overhead from the state-of-the-art compilation. 

The SAB compiler still cannot compete with our co-designed approach even if it targets a much denser architecture.
Grid17Q employs more connections, of course at the cost of $8\times$ lower yield rate compared to our XTree17Q.
However, even then the CNOT overhead for MtR on XTree17Q is only about $2.3\%$ of SAB on Grid17Q in most cases.

\textbf{Locality improvement:}
Analysis of mapping overheads shows that our ansatz construction improves gate locality.
At 10\% ansatz compression ratio, MtR on XTree17Q does not require any additional CNOTs most of the time.
We also observe that this mapping overhead jumps much faster from 70\% to 90\% compared to other gaps.
For example, the mapping overhead increases from 70\% to 90\% is about $2.9\times$ that of 50\% to 70\%, while the original CNOT count increment from 70\% to 90\% is only about $0.47\times$ that of 50\% to 70\%.
This is because our ansatz construction will first select \psc s with more gate locality so that they can be synthesized and mapped to XTree17Q efficiently.
But at compression ratios close to 1 (i.e. little compression), \psc s with poor locality will also be included in the ansatz, which makes the mapping overhead grow faster.

\section{Discussion and Future Directions}

In this paper, we advance variational quantum computational chemistry through a holistic software-hardware co-optimization from the algorithm, compiler, and hardware levels, outperforming conventional setups with significant benefits of multiple aspects.
This is the first attempt, to the best of our knowledge, that leverages the high-level application domain knowledge to coordinate the optimizations throughout the three levels from software to hardware in quantum computing.
Also our software-hardware co-optimization  is  not  targeting  a  particular  program  instance  and can broadly accommodate  the  full  family  of  computational  chemistry problems  with  such  structure.  
We believe that the co-optimization principle can also be applied to other promising application domains and hardware implementation technologies to boost the development of quantum computing.
Several further research directions are briefly discussed as follows:

\textbf{More physical systems:} 
This paper focused on chemical systems and the results can guide the development of new useful compounds.
Many other physical systems are also worth simulating.
For example, the Hubbard model~\cite{hubbard1963electron} in condensed matter physics can explain the transition between conducting and insulating systems.
These models may have different characteristics compared to a chemical system, e.g., periodic potential vs atomic potential, fermion vs boson. 
We expect that the Pauli-string-centric principle will still be applicable since the mathematics about simulating a Hamiltonian is invariant.
But the actual optimizations may need to change according to the characteristics of these models.


\textbf{Hardware architecture variants:} 
This paper focuses on the tree architecture with a minimized number of connections for a higher yield rate. 
However, it is not yet known how to find other Pareto-optimal designs.
We may also need to change the number of connections per qubit when scaling up and to improve CNOT fidelity. It can be interesting to consider tree structures with different degrees at different levels.
Moreover, for other hardware like ion traps, the main constraints can be different, and it is worth exploring how to extend the Pauli-string-centric principle to optimize quantum computational chemistry on other platforms.

\textbf{Deeper compiler optimization:} The compiler optimization in this paper is for the circuit synthesis and qubit mapping passes, which are essential in compiling a program to an executable circuit on a superconducting quantum processor.
Deeper compiler optimization is possible in at least two directions. 
First, other passes in the traditional compilation flow, e.g., gate cancellation~\cite{nam2018automated}, may be customized to variational quantum chemistry simulation programs.
Second, the variational quantum simulation is a numerical optimization algorithm.
It is thus possible to allow approximate compilation for more aggressive compiler optimization. 
Third, compiler-based error mitigation techniques~\cite{tirthak2020veritas, tannu2019ensemble, tannu2019mitigating} can also be incorporated to further reduce the simulation error.

\section{Related Work}

The techniques in this paper range across the algorithm, hardware, and compiler for variational quantum computational chemistry.
We briefly introduce related work for each of them.

\subsection{Algorithmic Optimization}
The major component in the VQE circuit is the parameterized ansatz.
UCCSD~\cite{peruzzo2014variational} is the ``standard" chemistry-inspired ansatz, but has a large size.
There have been several optimizations to reduce its size~\cite{lee2018generalized,grimsley2019adaptive,dallaire2019low, ryabinkin2018qubit, ryabinkin2020iterative,tang2019qubit}, but without considering specific hardware mapping overheads.
At the other extreme, ``hardware-efficient" ansatzes~\cite{kandala2017hardware} have been proposed which only employ gates that are easy to implement on the underlying hardware. However, these ansatzes
are unlikely to support large molecules since they do not consider any information about the chemical system to be simulated~\cite{mcclean2018barren,mcardle2020quantum}. Alternatively, several ansatz selection techniques rely on classical simulation of the molecule, and it is unclear how they scale to super-classical regimes~\cite{nam2020ground,eddins2021doubling}.
In contrast, the algorithm optimization proposed in this paper exploits information about the target system through \ps~comparison, and can maintain simulation accuracy as well as reduce hardware mapping overhead, and does not require classical simulation.

Additionally, there is prior work on optimizing the number of measurements required to evaluate the energy~\cite{measurement_reduction0, measurement_reduction1, measurement_reduction2, measurement_reduction3, measurement_reduction4}.
This type of optimization reduces the number of iterations of the inner loop in Figure~\ref{fig:vqeexample} and is orthogonal to our techniques which reduce the number of iterations in the outer loop as well as the size of the circuit itself.
These optimizations can be employed together with our techniques.

\subsection{Compiler Optimization}

A large body of work exists on mapping quantum circuits to hardware~\cite{murali2019noise,li2019tackling,zulehner2018efficient,tannu2019not,childs2019circuit}. These algorithms are invoked after a quantum circuit is already synthesized and are general-purpose with little assumption regarding the input programs or the underlying hardware architectures. 

High-level semantics have recently been considered in compiler optimizations.
Cowtan et al. recently proposed a method for compiling UCC ansatzes by partitioning \ps s into sets, but not considering the underlying architecture~\cite{cowtan2020generic}.
An architecture-aware synthesis for phase-polynomial quantum circuits was proposed in~\cite{de2020architecture}.

In contrast, this paper uses the \psc~to devise a new compilation flow based not only on the chemistry simulation domain knowledge but also on the underlying architecture.
Starting from a \ps~IR, it achieves unprecedented mapping overhead reduction by combining synthesis and mapping in a single pass.

\subsection{Application-specific Quantum Processor Architecture}

An application-specific quantum architecture was proposed by
Wilhelm et al. for a specific Fermi-Hubbard model simulation, based on a superconducting planar architecture~\cite{dallaire2016quantum,liebermann2017implementation}.
Recently, an end-to-end design flow has also been proposed to generate optimized superconducting quantum processor architectures for different individual quantum programs~\cite{li2020towards}.
These architectures are circuit-specific rather than domain-specific, as they exploit low-level gate patterns but not high-level domain knowledge and do not generalize to families of circuits.
For trapped ion technology,~\cite{brown2016co} provided a forward-looking overview of co-designing trapped ion machines. 
Murali et al. also proposed a toolflow to evaluate the architecture design of trapped ion quantum computers over a benchmark suite~\cite{murali2020architecting}.
Our architecture design, which integrates the algorithm-level domain knowledge, is a concrete optimized design with compiler support to accommodate various variational quantum chemistry programs with different simulation targets.



\section{Conclusion}

In this paper, we advance variational quantum chemistry simulation through a holistic software-hardware co-optimization at the algorithm, compiler, and hardware levels.
We show that variational quantum chemistry programs can be significantly simplified without complex derivative calculation, and they can be efficiently mapped onto a high yield superconducting quantum processor with very sparse connections.
The three proposed optimizations can accommodate simulating various chemical systems and  bring a wide range of advantages from software to hardware.
The design principle and the results from this paper could guide future development of quantum software and hardware infrastructures.

\section*{Acknowledgements}
We thank the anonymous reviewers for the constructive comments. We thank Moein Malekakhlagh for helpful discussions and Hiroshi Horii for help with experiment setup. G. L. was in part funded by NSF QISE-NET fellowship under the award DMR-1747426.

\bibliographystyle{unsrt}
\bibliography{main}

\end{document}